\documentclass[twocolumn,english,showpacs]{revtex4}
\usepackage{amsmath,amssymb,dcolumn}
\usepackage[dvips]{graphics}
\usepackage{graphicx}
\def\imo{i}
\begin{document}
\title{(In)stability of D-dimensional black holes in Gauss-Bonnet theory}
\author{R. A. Konoplya}\email{konoplya@tap.scphys.kyoto-u.ac.jp}
\affiliation{Department of Physics, Kyoto University, Kyoto 606-8501, Japan}
\author{A. Zhidenko}\email{zhidenko@fma.if.usp.br}
\affiliation{Instituto de F\'{\i}sica, Universidade de S\~{a}o Paulo \\
C.P. 66318, 05315-970, S\~{a}o Paulo-SP, Brazil}

\begin{abstract}
We make an extensive study of evolution of gravitational
perturbations of D-dimensional black holes in Gauss-Bonnet theory.
There is an instability at higher multi-poles $\ell$ and large
Gauss-Bonnet coupling $\alpha$ for $D= 5, 6$, which is stabilized at
higher $D$. Although small negative gap of the effective potential for
scalar type of gravitational perturbations, exists for
higher $D$ and whatever $\alpha$, it does not lead to any instability.
\end{abstract}

\pacs{04.30.Nk,04.50.+h}
\maketitle

\section{Introduction}

In recent years, higher dimensional black holes have been in the focus
of high energy physics research. They are essential for our
understanding of key moments of string theory, quantum gravity and
brane-world scenarios. The most important classical property of
black holes is evidently stability: unstable black holes simply
cannot exist in our world. In addition, when considering higher
dimensional black holes in the context of  anti-de Sitter/Conformal
Field Theory (AdS/CFT) correspondence, that is as a dual backgrounds
rather than as real black holes, instability means a phase
transition in the dual theory \cite{Gubser} being essential
for understanding the field theory at finite temperature and high
coupling. An opportunity of creating of mini black holes at particle
collisions in Large Hadron Collider, according to Tev gravity extra
dimensional scenarios, also gave a strong impetus to extensive
studies of different properties of higher dimensional black holes,
including quasinormal modes \cite{QNMshigherD} and Hawking radiation
\cite{Hawking_higherD}.

The study of evolution of gravitational perturbations of the
D-dimensional black holes in pure Einstein theory started from the
work of Ishibashi and Kodama \cite{Kodama} who managed to reduce the
cumbersome perturbation equations to the usual Schrodinger-like
form. Yet the resulting effective potential of the wave equation is
not always positive definite, so that stability is not taken for granted.
In a few cases the technique of the so-called S-deformations,
i.e. of deformations of the wave equation which does not touch
stability properties, helped to transform a potential to a positive
definite form, thereby proving the stability. In a general case, an
extensive numerical investigation of evolution of gravitational
perturbations allowed to prove stability for higher dimensional
Reissner-Nordstr\"om-de Sitter black holes with arbitrary charge and
$\Lambda$-term \cite{MyNPB1}.

At the same time, the quantum gravity corrections to classical
general relativity implies the existence of the so-called
Gauss-Bonnet term in the dominant order correction, that is, the
term in the Lagrangian squared in curvature. This term vanishes when
$D=4$. Therefore, the black holes in the Einstein-Gauss-Bonnet
theory \cite{Boulware:1985wk} have attracted considerable interest recent
years \cite{GBgeneral}. In particular, scalar field quasinormal
modes of asymptotically flat Gauss-Bonnet black holes
\cite{GBQNMsmy1} and of asymptotically dS/AdS Gauss-Bonnet black
holes were considered in \cite{GBQNMsmy2}. The scalar quasinormal
modes of Gauss-Bonnet black holes in the regime of high damping were
considered in \cite{GBQNMs_asympt}. The scalar field, propagating in
the background of a black hole, although gives the qualitative
picture of evolution of perturbations, is not responsible for
stability, so that perturbations of Einstein-Gauss-Bonnet equations
must be considered instead \cite{Dotti1}, \cite{Dotti2}. Thus, Dotti
and Gleiser reduced the Einstein-Gauss-Bonnet perturbed equations to
a wave like form with some effective potentials. They found an
instability for the two particular cases: scalar type of
gravitational perturbations for $D=5$, and tensor type of
gravitational perturbations for $D=6$, leaving the analysis of
stability of general $D$ open. The problem was that analytical
treatment of stability is difficult in many cases: even if we
have a wave like equation with a potential (extremely cumbersome in
the Gauss-Bonnet case), negative gaps in potentials cannot be easily
removed by the S-deformations, because one needs to know an ansatz
that transforms the potential to a positive definite one.

The aim of our work is to perform a complete numerical
analysis of the evolution of gravitational perturbations for
D-dimensional Gauss-Bonnet black holes with $D = 5-11$, what is
motivated by string theory and quantum gravity, and to determine the
stability and instability regions for these black holes.

The paper is organized as follows: Sec. II introduces the metric and
wave like equations for Gauss-Bonnet (GB) black holes. Sec. III
describes the methods of time domain integration used here and shows
all numerical data for evolution  of perturbations and quasinormal
modes. Sec IV discusses the results obtained.

\begin{widetext}
\section{The perturbation wave equations for Gauss-Bonnet black holes}

The Lagrangian of the Einstein-Gauss-Bonnet action is
\begin{equation}
I = \frac{1}{16 \pi G_{D}} \int{d^{D} x \sqrt{-g} R} + \alpha^\prime
\int d^{D} x \sqrt{-g} (R_{abcd} R^{abcd}- 4 R_{cd}R^{cd} + R^2).
\end{equation}
Here $\alpha^\prime$ is a positive coupling constant.

The metric has the form,
\begin{equation}\label{metric}
ds^2=f(r)dt^2-\frac{dr^2}{f(r)}-r^2d\Omega_{D-2}^2,
\end{equation}
\begin{equation}\nonumber
f(r)=1+\frac{r^2}{\alpha(D-3)(D-4)}\left(1-q(r)\right), \qquad
q(r)=\sqrt{1+\frac{4\alpha(D-3)(D-4)\mu}{(D-2)r^{D-1}}},
\end{equation}
where $\alpha = 16 \pi G_{D} \alpha^\prime$.

In order to measure all the quantities in terms of the black hole horizon $r_0$ radius we parameterize
the black hole mass as
\begin{equation}
\mu=\frac{(D-2)r_0^{D-3}}{4}\left(2+\frac{\alpha(D-3)(D-4)}{r_0^2}\right).
\end{equation}

As was shown in \cite{Dotti1}, \cite{Dotti2}, the gravitational
perturbations of a Gauss-Bonnet black hole can be decoupled from
their angular part and reduced to the wave-like equation of the form
\begin{equation}\label{wave-like}
\left(\frac{\partial^2}{\partial t^2}-\frac{\partial^2}{\partial r_\star^2} + V(r)\right)\Psi(t,r) = 0, \qquad dr_\star=\frac{dr}{f(r)}.
\end{equation}
with the effective potentials which have very cumbersome form
\cite{Dotti1}, \cite{Dotti2}. After some algebra, we managed to
simplify the potentials for the tensor, vector and scalar types of
the gravitational perturbation respectively:

\begin{eqnarray}
V_t(r)&=&f(r)\frac{\lambda}{r^2}\left(3-\frac{B(r)}{A(r)}\right)+\frac{1}{\sqrt{r^{D-2}A(r)q(r)}}\frac{d^2}{dr_\star^2}\sqrt{r^{D-2}A(r)q(r)},\\
V_v(r)&=&f(r)\frac{(D-2)c}{r^2}A(r)+\sqrt{r^{D-2}A(r)q(r)}\frac{d^2}{dr_\star^2}\frac{1}{\sqrt{r^{D-2}A(r)q(r)}},\\
V_s(r)&=&\frac{f(r)U(r)}{64 r^2(D-3)^2A(r)^2q(r)^8(4c q(r) + (D-1)R (q(r)^2-1))^2},\\\nonumber
\end{eqnarray}

We used the following dimensionless quantities
\begin{eqnarray}\nonumber
A(r)&=&\frac{1}{q(r)^2}\left(\frac{1}{2}+\frac{1}{D-3}\right)+\left(\frac{1}{2}-\frac{1}{D-3}\right),\\\nonumber
B(r)&=&A(r)^2\left(1+\frac{1}{D-4}\right)+\left(1-\frac{1}{D-4}\right),\\\nonumber
R&=&\frac{r^2}{\alpha(D-3)(D-4)},
\end{eqnarray}
\begin{eqnarray}\nonumber
U(r)&=&5 (D - 1)^6 R^2 (1 + R) - 3 (D - 1)^5 R ((D - 1) R^2 + 24 c (1 + R)) q(r) + \\\nonumber&& +
 2 (D - 1)^4 (24 c (D - 1) R^2 +
    168 c^2 (1 + R) - (D - 1) R^2 (-3 + 5 R + 7 D (1 + R))) q(r)^2 + \\\nonumber&& +
 2 (D - 1)^4 R (-184 c^2 + (D - 1) (13 + D) R^2 +
    c (-84 + 44 R + 84 D (1 + R))) q(r)^3 + \\\nonumber&& +
    (D - 1)^3 (384 c^3 - 48 c (2 + D (3 D - 5)) R^2 +
    192 c^2 (-11 + D + (-15 + D) R) + \\\nonumber&& + (D - 1) R^2 (-3 (7 + 55 R) +
       D (26 + 106 R + 7 D (1 + R)))) q(r)^4 + \\\nonumber&& +
       (D - 1)^3 R (-64 c^2 (D - 38) + (D - 1) (71 + D (7 D - 90)) R^2 + \\\nonumber&& +
    16 c (303 + 255 R + 13 D^2 (1 + R) - 2 D (73 + 81 R))) q(r)^5 + \\\nonumber&& +
 4 (D - 1)^2 (96 c^3 (-7 + D) -
    8 c (D - 1) (145 - 74 D + 6 D^2) R^2 - \\\nonumber&& -
    8 c^2 (9 - 175 R + D (-58 - 34 R + 11 D (1 + R))) + (D - 1) R^2 (-5 (79 + 23 R) + \\\nonumber&& +
       D (5 (57 + 41 R) + D (-81 - 89 R + 7 D (1 + R))))) q(r)^6 - \\\nonumber&& -
 4 (D - 1)^2 R (8 c^2 (43 + (72 - 13 D) D) + (D - 1) (-63 + D (99 + D (-49 + 5 D))) R^2 + \\\nonumber&& +
    4 c (321 + 465 R + D (121 - 39 R + D (-123 - 107 R + 17 D (1 + R))))) q(r)^7 + \\\nonumber&& +
       (D - 1) (128 c^3 (-9 + D) (D - 5) + 32 c (D - 1) (246 + D (9 + D (-55 + 8 D))) R^2 + \\\nonumber&& +
    64 c^2 (D - 5) (D^2 - 3 + (49 + (D - 4) D) R) - \\\nonumber&& -
    (D - 1) R^2 (1173 + 565 R + D (-4 (997 + 349 R) + D (6 (393 + 217 R) + D (-548 - 452 R + 45 D (1 + R)))))) q(r)^8 + \\\nonumber&& +
    (D - 1) R (-64 c^2 (D - 5) (36 + D (-13 + 3 D)) + (D - 1) (635 + D (-1204 + 3 D (294 + D (-92 + 9 D)))) R^2 - \\\nonumber&& -
    8 c (D - 5) (63 + 31 R + D (127 + 191 R + D (-47 + D + (-79 + D) R)))) q(r)^9 + \\\nonumber&& +
 2 (D - 5) (64 c^3 (D - 5) (D - 3) + 8 c (D - 1) (-27 + D (141 + (-43 + D) D)) R^2 + \\\nonumber&& +
    8 c^2 (D - 5) (-3 + 77 R + D (D - 2 + (D - 18) R)) + (D - 1)^2 R^2 (-33 (R - 7) + \\\nonumber&& +
       D (59 + 43 R + D (-59 - 35 R + 9 D (1 + R))))) q(r)^{10} - \\\nonumber&& -
 2 (D - 5) R (24 c^2 (-11 + D) (D - 5) (D - 3) + (D - 1)^2 (-65 + D (81 + D (7 D - 39))) R^2 + \\\nonumber&& +
    12 c (-7 + D) (D - 5) (D - 3) (D - 1) (1 + R)) q(r)^{11} + \\\nonumber&& +
    (D - 5)^2 (-1 + D) R^2 (16 c (26 + (D - 9) D) + (D - 1) (77 - 3 R + D (-18 + D + (D - 2) R))) q(r)^{12} + \\\nonumber&& +
    (D - 5)^2 (D - 3)^2 (D - 1)^2 R^3 q(r)^{13},
\end{eqnarray}
$\lambda=(D-2)(c+1)=\ell(\ell + D - 3)$ is the eigenvalue of the angular
part of the Laplacian.


\section{The evolution of perturbations in time domain}

We study the ringing of GB black hole using a numerical
characteristic integration method \cite{Price-Pullin}, that uses the
light-cone variables $u = t - r_\star$ and $v = t + r_\star$. In the
characteristic initial value problem, initial data are specified on
the two null surfaces $u = u_{0}$ and $v = v_{0}$. The
discretization scheme we used, is
\begin{equation}\label{d-uv-eq}
\Psi(N) = \Psi(W) + \Psi(E) - \Psi(S) -\Delta^2\frac{V(W)\Psi(W) + V(E)\Psi(E)}{8} + \mathcal{O}(\Delta^4) \ ,
\end{equation}
where we have used the following definitions for the points: $N =(u + \Delta, v + \Delta)$, $W = (u + \Delta, v)$, $E = (u, v + \Delta)$ and $S = (u,v)$.
\end{widetext}

To see the correct time-domain profile at late time we need to calculate precisely the values of the effective potential which are used in (\ref{d-uv-eq}). In order to do this we must integrate numerically the equation for the tortoise coordinate and then solve it with respect to the radial coordinate with high accuracy (we worked with a precision of $\sim 2^{-90}$). We used simple Runge-Kutta method for the integration. Since it interpolates the function by a cubic spline at each step we are able to find analytically $r(r_\star)$ at each step. The final \texttt{C++} programm that finds the time-domain profiles with arbitrary precision is available from the last author upon request.

\begin{figure*}
\begin{tabular}{rl}
\begin{minipage}{.15\textwidth}
\begin{tabular}{|r|c|}
\hline
$\ell$&$\alpha$\\
\hline
$8$&$1.346$\\
$10$&$1.274$\\
$12$&$1.227$\\
$16$&$1.170$\\
$20$&$1.136$\\
$32$&$1.087$\\
$40$&$1.070$\\
$50$&$1.058$\\
$64$&$1.047$\\
\hline
\end{tabular}
\end{minipage}
&
\begin{minipage}{.7\textwidth}
\includegraphics[width=\textwidth,clip]{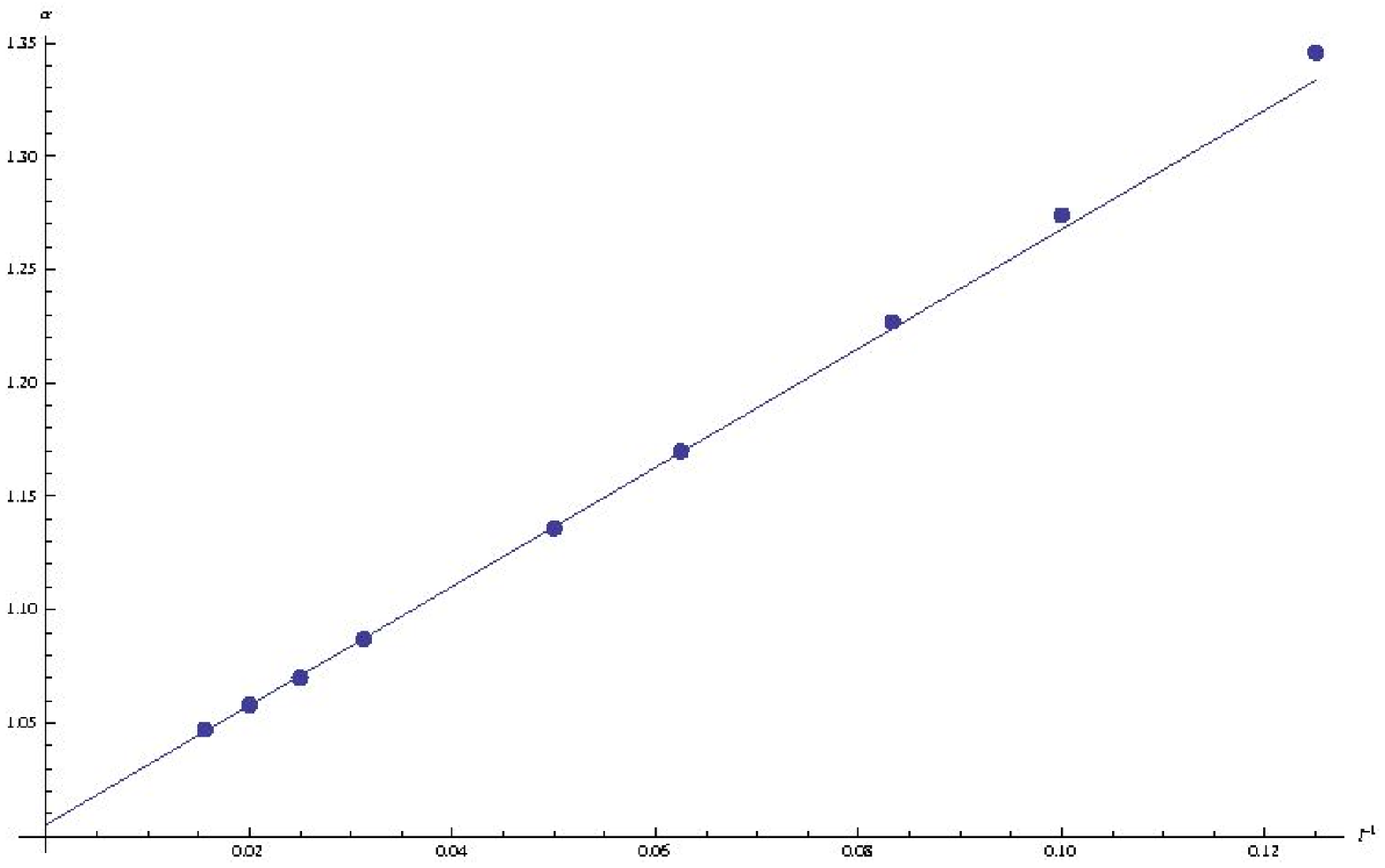}
\end{minipage}\\
\end{tabular}
\caption{Threshold $\alpha$ as a function of the inverse multipole
number $\ell$. Tensor type of gravitational perturbations $D=6$. \\The
points $\ell=16,20,32,40,50,64$ were fit by the line $\alpha =
2.627\ell^{-1}+1.005$. The theoretical result is
$\alpha_t\approx1.006$.}
\end{figure*}

\begin{figure*}
\begin{tabular}{rl}
\begin{minipage}{.15\textwidth}
\begin{tabular}{|r|c|}
\hline
$\ell$&$\alpha$\\
\hline
$8$&$0.268$\\
$10$&$0.258$\\
$12$&$0.250$\\
$16$&$0.241$\\
$20$&$0.235$\\
$32$&$0.225$\\
$40$&$0.222$\\
$50$&$0.219$\\
\hline
\end{tabular}
\end{minipage}
&
\begin{minipage}{.7\textwidth}
\includegraphics[width=\textwidth,clip]{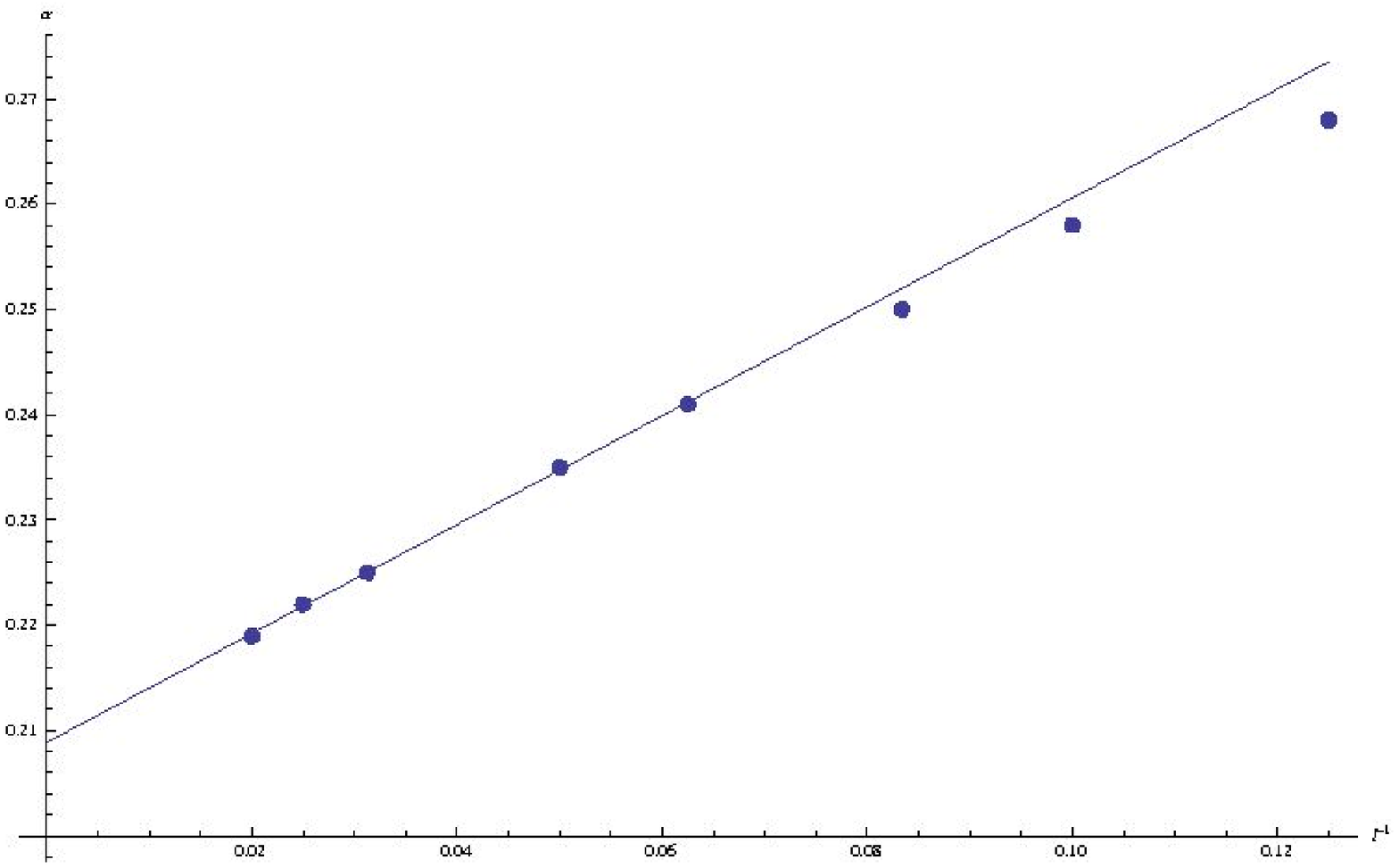}
\end{minipage}\\
\end{tabular}
\caption{Threshold $\alpha$ as a function of the inverse multipole
number $\ell$. Scalar type of gravitational perturbations $D=5$. \\The
points $\ell=16,20,32,40,50$ were fit by the line $\alpha =
0.517\ell^{-1}+0.209$. The theoretical result is
$\alpha_t\approx0.207$.}
\end{figure*}

\begin{table*}
\caption{Fundamental QNMs for GB black hole perturbation of tensor type ($\ell=2$). $\alpha$ and frequencies are measured in units of the horizon radius.}
\begin{tabular}{|c|c|c|c|c|c|c|c|}
\hline
$\alpha$&$D=5$&$D=6$&$D=7$&$D=8$&$D=9$&$D=10$&$D=11$\\
\hline
$0.1$&$1.5898-0.3406\imo$&$2.0112-0.4521\imo$&$2.3697-0.5541\imo$&$2.7006-0.6492\imo$&$3.0212-0.7342\imo$&$3.3386-0.8092\imo$&$3.6565-0.8753\imo$\\
$0.2$&$1.6435-0.3289\imo$&$1.9557-0.4269\imo$&$2.2166-0.5246\imo$&$2.4761-0.6105\imo$&$2.7428-0.6827\imo$&$3.0167-0.7429\imo$&$3.2970-0.7936\imo$\\
$0.3$&$1.6766-0.3157\imo$&$1.8881-0.4086\imo$&$2.0901-0.5034\imo$&$2.3143-0.5820\imo$&$2.5541-0.6456\imo$&$2.8045-0.6958\imo$&$3.0633-0.7340\imo$\\
$0.4$&$1.6946-0.3014\imo$&$1.8217-0.3942\imo$&$1.9874-0.4862\imo$&$2.1907-0.5599\imo$&$2.4126-0.6175\imo$&$2.6448-0.6581\imo$&$2.8886-0.6816\imo$\\
$0.5$&$1.7017-0.2870\imo$&$1.7603-0.3823\imo$&$1.9024-0.4718\imo$&$2.0919-0.5423\imo$&$2.2990-0.5953\imo$&$2.5154-0.6242\imo$&$2.7514-0.6299\imo$\\
$0.6$&$1.7012-0.2731\imo$&$1.7046-0.3720\imo$&$1.8308-0.4594\imo$&$2.0100-0.5280\imo$&$2.2031-0.5771\imo$&$2.4066-0.5902\imo$&$2.6438-0.5788\imo$\\
$0.7$&$1.6956-0.2600\imo$&$1.6541-0.3630\imo$&$1.7693-0.4486\imo$&$1.9403-0.5163\imo$&$2.1190-0.5612\imo$&$2.3149-0.5544\imo$&$2.5596-0.5312\imo$\\
$0.8$&$1.6863-0.2478\imo$&$1.6083-0.3549\imo$&$1.7158-0.4390\imo$&$1.8799-0.5068\imo$&$2.0427-0.5461\imo$&$2.2389-0.5176\imo$&$2.4927-0.4894\imo$\\
$0.9$&$1.6745-0.2366\imo$&$1.5667-0.3476\imo$&$1.6685-0.4303\imo$&$1.8265-0.4991\imo$&$1.9716-0.5299\imo$&$2.1767-0.4821\imo$&$2.4378-0.4532\imo$\\
$1.0$&$1.6611-0.2262\imo$&$1.5286-0.3409\imo$&$1.6264-0.4225\imo$&$1.7790-0.4931\imo$&$1.9054-0.5095\imo$&$2.1255-0.4497\imo$&$2.3915-0.4219\imo$\\
\hline
\end{tabular}
\end{table*}

\begin{table*}
\caption{Fundamental QNMs for GB black hole perturbation of vector
type ($\ell=2$). $\alpha$ and frequencies are measured in units of the
horizon radius.}
\begin{tabular}{|c|c|c|c|c|c|c|c|}
\hline
$\alpha$&$D=5$&$D=6$&$D=7$&$D=8$&$D=9$&$D=10$&$D=11$\\
\hline
$0.1$&$1.0887-0.3196\imo$&$1.4227-0.4372\imo$&$1.7460-0.5373\imo$&$2.0608-0.6281\imo$&$2.3712-0.7092\imo$&$2.6805-0.7815\imo$&$2.9919-0.8450\imo$\\
$0.2$&$1.0490-0.3077\imo$&$1.3321-0.4063\imo$&$1.6016-0.4964\imo$&$1.8664-0.5768\imo$&$2.1330-0.6447\imo$&$2.4036-0.7006\imo$&$2.6800-0.7475\imo$\\
$0.3$&$1.0128-0.2950\imo$&$1.2582-0.3831\imo$&$1.4947-0.4682\imo$&$1.7315-0.5406\imo$&$1.9735-0.5975\imo$&$2.2239-0.6401\imo$&$2.4836-0.6727\imo$\\
$0.4$&$0.9797-0.2829\imo$&$1.1973-0.3654\imo$&$1.4113-0.4466\imo$&$1.6287-0.5123\imo$&$1.8540-0.5583\imo$&$2.0923-0.5879\imo$&$2.3448-0.6081\imo$\\
$0.5$&$0.9494-0.2719\imo$&$1.1459-0.3510\imo$&$1.3434-0.4293\imo$&$1.5454-0.4886\imo$&$1.7589-0.5229\imo$&$1.9919-0.5400\imo$&$2.2426-0.5515\imo$\\
$0.6$&$0.9217-0.2621\imo$&$1.1014-0.3390\imo$&$1.2862-0.4151\imo$&$1.4752-0.4676\imo$&$1.6812-0.4893\imo$&$1.9140-0.4965\imo$&$2.1652-0.5029\imo$\\
$0.7$&$0.8961-0.2535\imo$&$1.0627-0.3287\imo$&$1.2369-0.4030\imo$&$1.4143-0.4479\imo$&$1.6175-0.4570\imo$&$1.8525-0.4580\imo$&$2.1045-0.4619\imo$\\
$0.8$&$0.8726-0.2458\imo$&$1.0286-0.3197\imo$&$1.1935-0.3927\imo$&$1.3606-0.4286\imo$&$1.5652-0.4266\imo$&$1.8030-0.4244\imo$&$2.0554-0.4272\imo$\\
$0.9$&$0.8508-0.2385\imo$&$0.9982-0.3116\imo$&$1.1547-0.3837\imo$&$1.3131-0.4089\imo$&$1.5220-0.3987\imo$&$1.7620-0.3953\imo$&$2.0144-0.3978\imo$\\
$1.0$&$0.8302-0.2318\imo$&$0.9712-0.3042\imo$&$1.1194-0.3758\imo$&$1.2713-0.3886\imo$&$1.4859-0.3735\imo$&$1.7275-0.3700\imo$&$1.9794-0.3725\imo$\\
\hline
\end{tabular}
\end{table*}

\begin{table*}
\caption{Fundamental QNMs for GB black hole perturbation of scalar type ($\ell=2$). $\alpha$ and frequencies are measured in units of the horizon radius.}
\begin{tabular}{|c|c|c|c|c|c|c|c|}
\hline
$\alpha$&$D=5$&$D=6$&$D=7$&$D=8$&$D=9$&$D=10$&$D=11$\\
\hline
$0.1$&$0.8248-0.2457\imo$&$0.9386-0.3557\imo$&$1.1592-0.5211\imo$&$0.6879-0.4592\imo$&$0.7935-0.4296\imo$&$0.8833-0.4417\imo$&$0.9555-0.4750\imo$\\
$0.2$&$0.7549-0.2566\imo$&$?$&$0.5723-0.3376\imo$&$0.7443-0.2782\imo$&$0.8703-0.2926\imo$&$0.9715-0.3382\imo$&$1.0501-0.4015\imo$\\
$0.3$&instability&$?$&$0.5939-0.2163\imo$&$0.7631-0.1970\imo$&$0.8947-0.2281\imo$&$1.0044-0.2862\imo$&$1.0905-0.3636\imo$\\
$0.4$&instability&$0.1906-0.2953\imo$&$0.6010-0.1533\imo$&$0.7665-0.1525\imo$&$0.9018-0.1893\imo$&$1.0190-0.2520\imo$&$1.1132-0.3374\imo$\\
$0.5$&instability&$0.2699-0.2067\imo$&$0.6015-0.1164\imo$&$0.7639-0.1244\imo$&$0.9020-0.1629\imo$&$1.0255-0.2268\imo$&$1.1277-0.3168\imo$\\
$0.6$&instability&$0.3061-0.1515\imo$&$0.5989-0.0925\imo$&$0.7588-0.1051\imo$&$0.8990-0.1435\imo$&$1.0278-0.2069\imo$&$1.1376-0.2994\imo$\\
$0.7$&instability&$0.3264-0.1153\imo$&$0.5948-0.0761\imo$&$0.7528-0.0910\imo$&$0.8946-0.1285\imo$&$1.0277-0.1906\imo$&$1.1445-0.2841\imo$\\
$0.8$&instability&$0.3387-0.0902\imo$&$0.5899-0.0642\imo$&$0.7463-0.0804\imo$&$0.8895-0.1166\imo$&$1.0261-0.1770\imo$&$1.1494-0.2704\imo$\\
$0.9$&instability&$0.3465-0.0722\imo$&$0.5847-0.0554\imo$&$0.7398-0.0720\imo$&$0.8840-0.1069\imo$&$1.0236-0.1653\imo$&$1.1527-0.2579\imo$\\
$1.0$&instability&$0.3513-0.0589\imo$&$0.5793-0.0486\imo$&$0.7334-0.0653\imo$&$0.8785-0.0988\imo$&$1.0205-0.1551\imo$&$1.1548-0.2464\imo$\\
\hline
\end{tabular}
\end{table*}
\begin{figure*}
\includegraphics[width=.5\textwidth,clip]{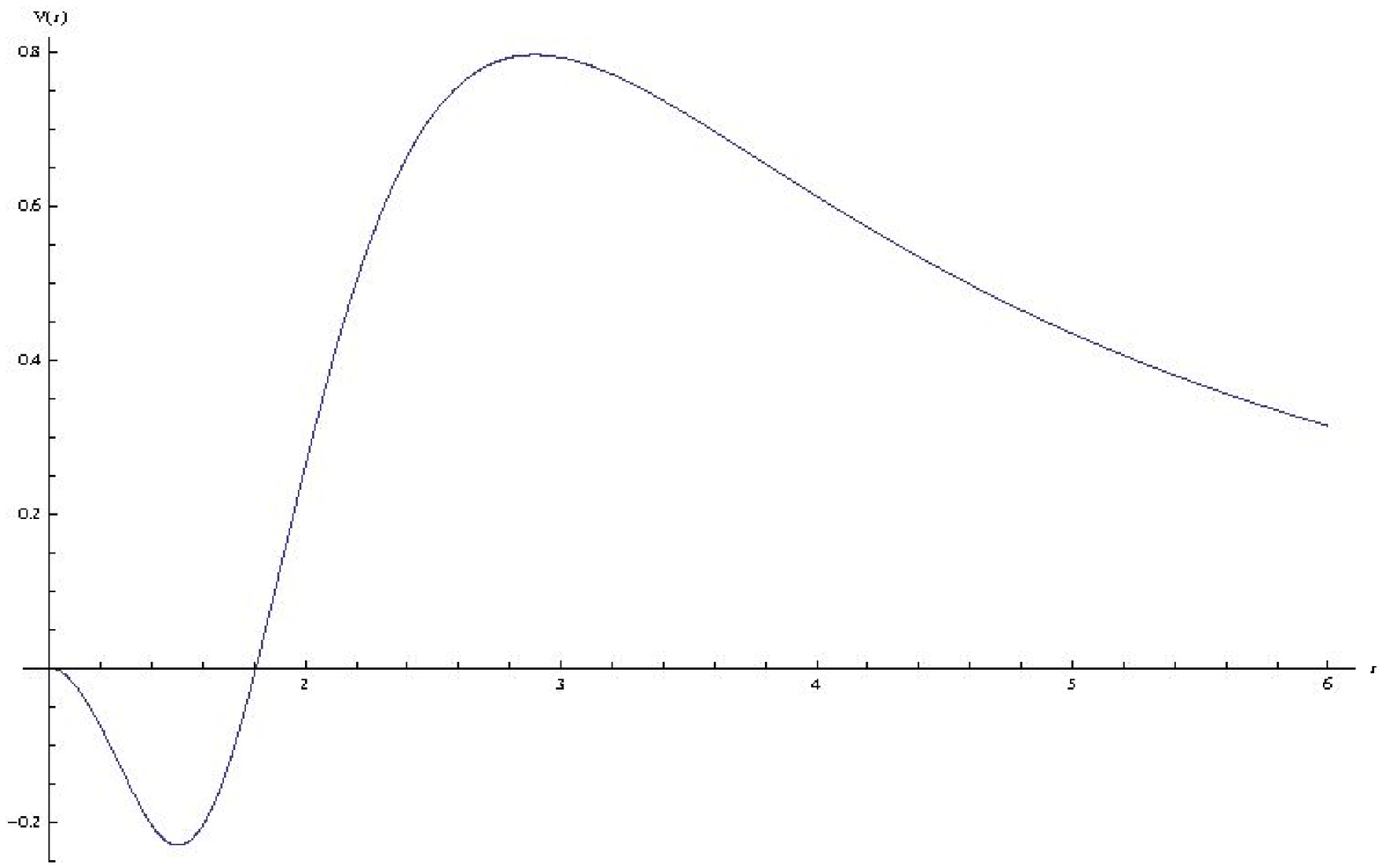}\includegraphics[width=.5\textwidth,clip]{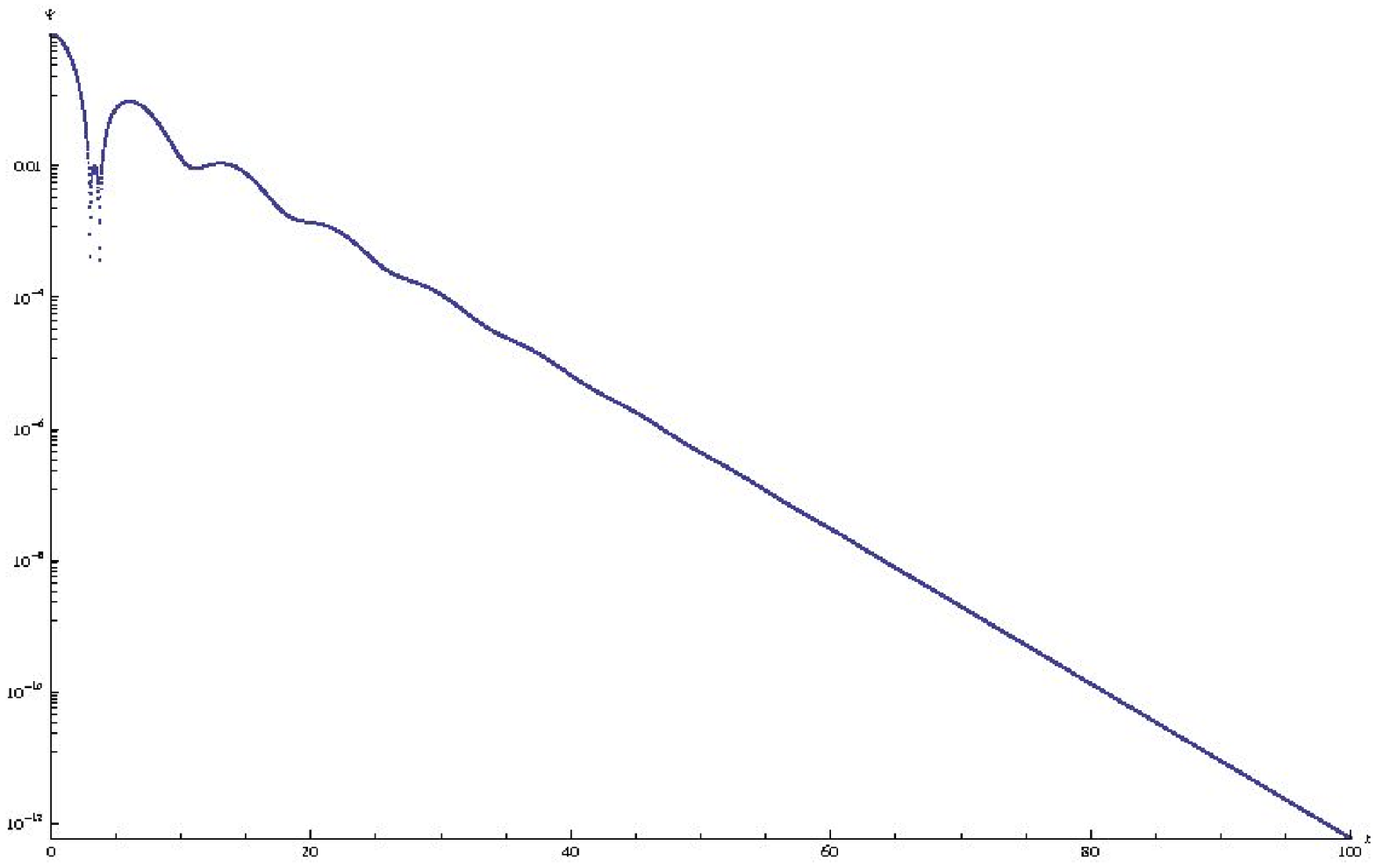}
\caption{Potential and profile for GB black hole perturbation of scalar type ($D=6$, $l=2$, $\alpha=0.3$). The negative gap does not lead to instability. It causes exponentially damping ``tails'' to appear just after the initial outburst. Therefore we are unable to see QN ringing.}
\end{figure*}

\begin{figure*}
\includegraphics[width=.3\textwidth,clip]{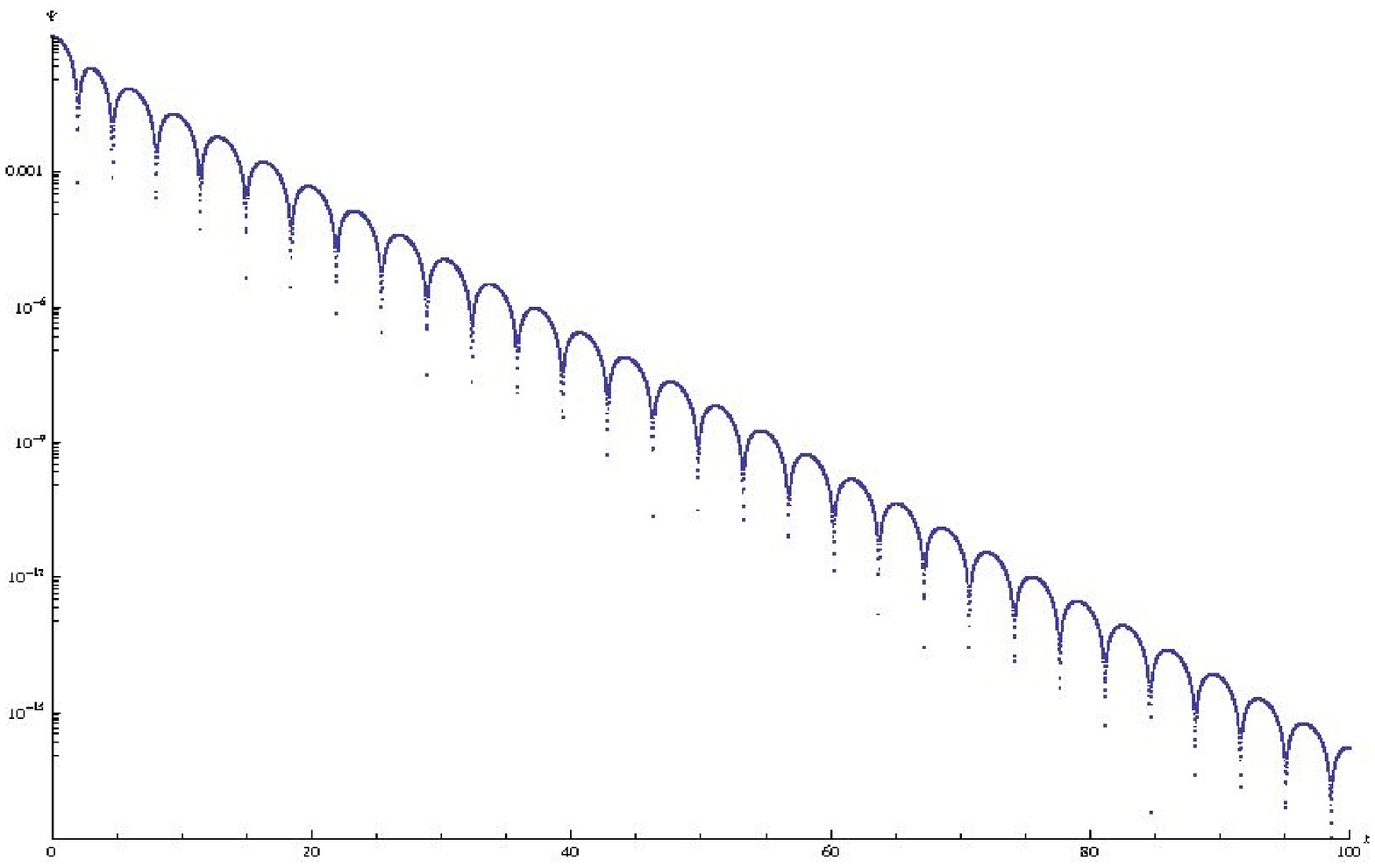}
\includegraphics[width=.3\textwidth,clip]{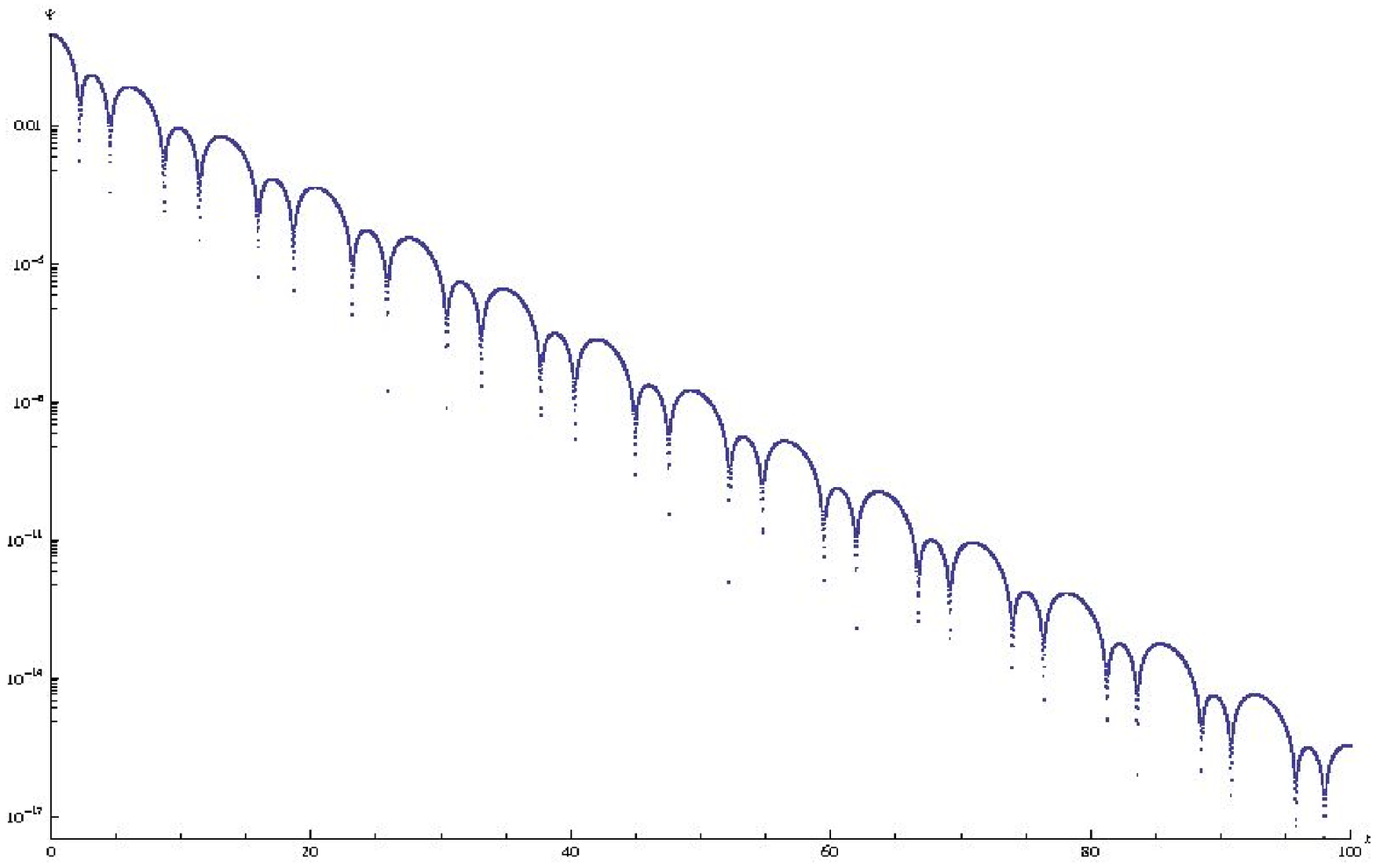}
\includegraphics[width=.3\textwidth,clip]{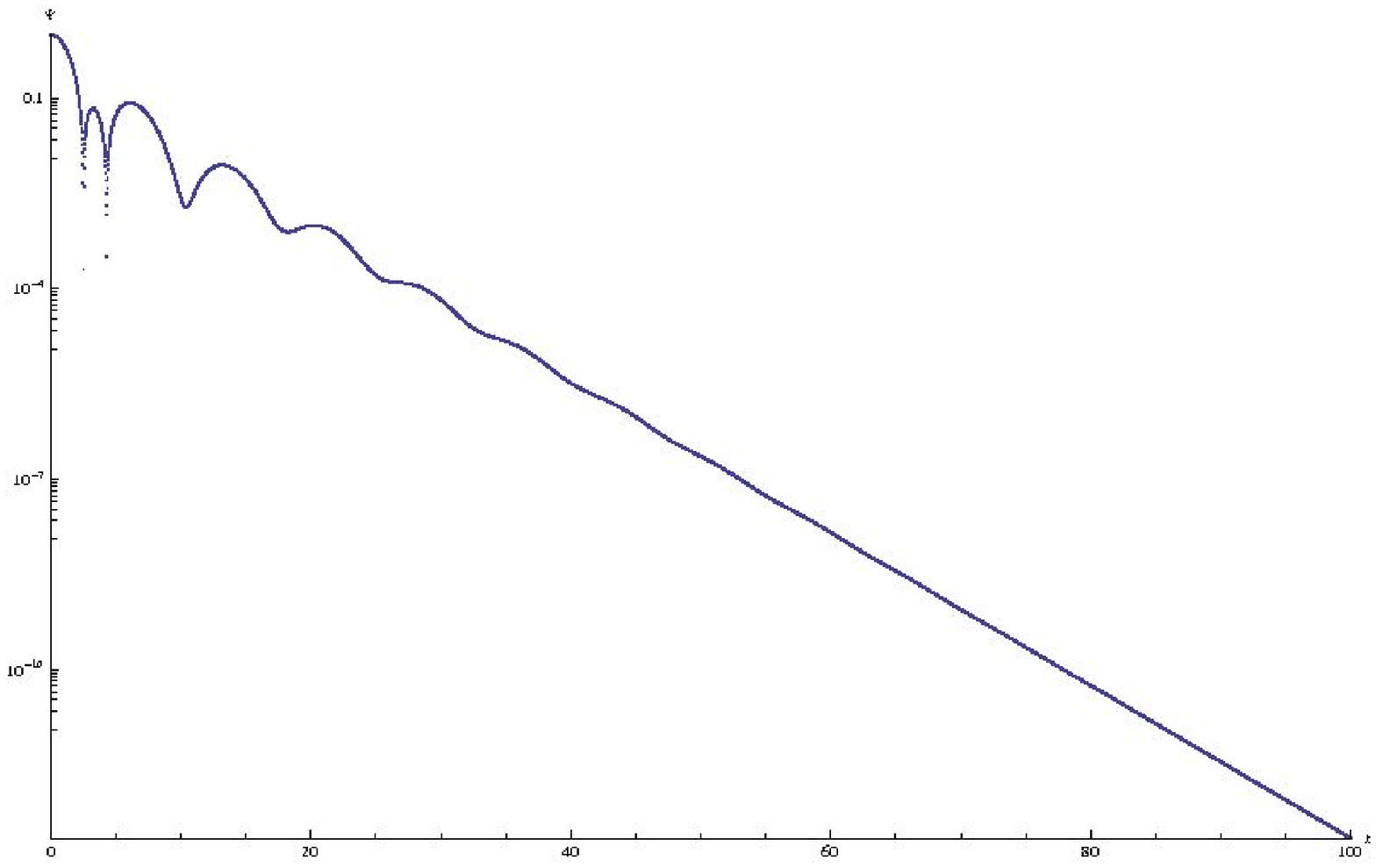}
\includegraphics[width=.3\textwidth,clip]{D=6.l=2.a=0.300.scalar.eps}
\includegraphics[width=.3\textwidth,clip]{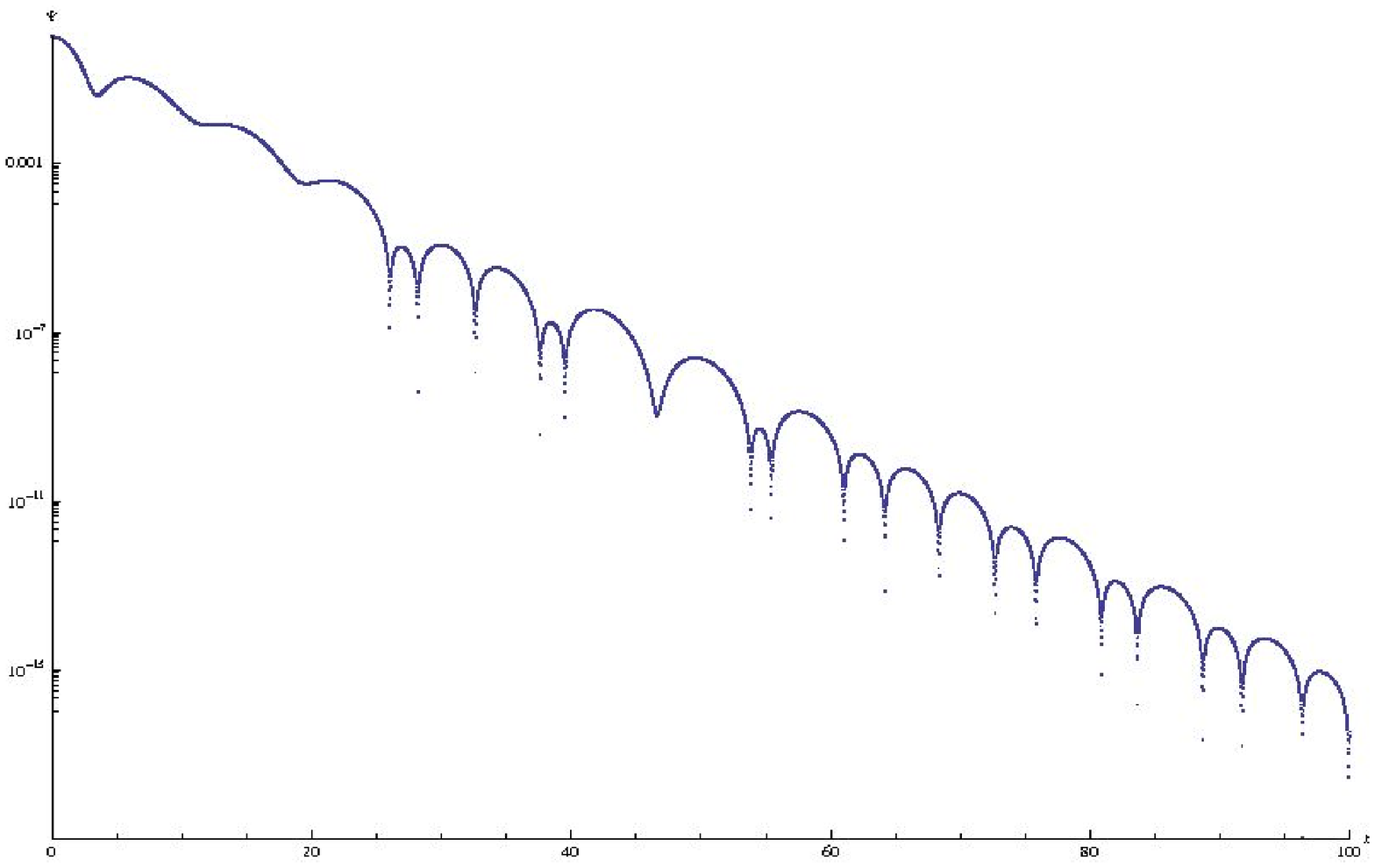}
\includegraphics[width=.3\textwidth,clip]{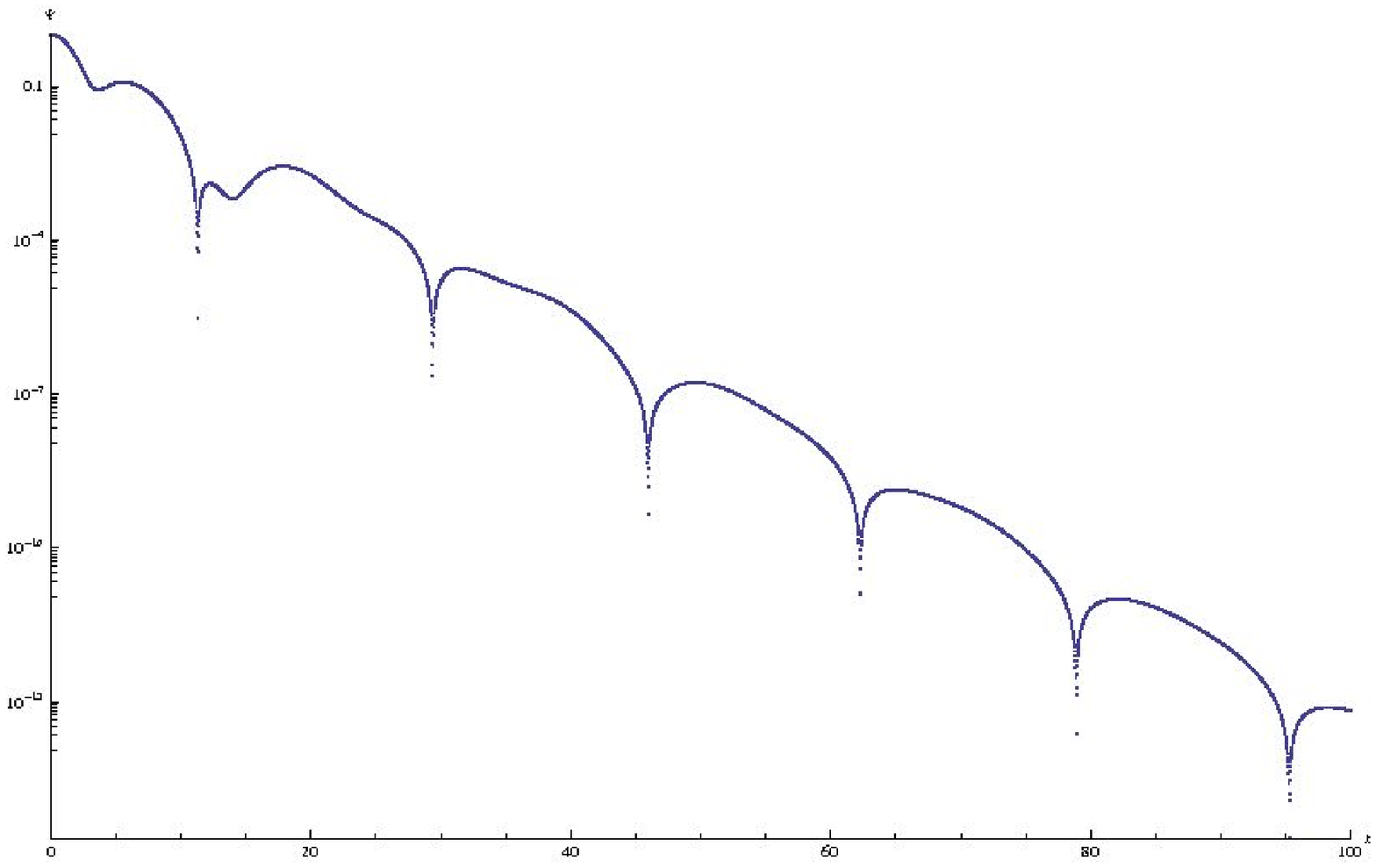}
\caption{Time-domain profiles for the ``region of the irregular QN-ringing" of the gravitational perturbation of scalar type $D=6$, $\ell=2$, $\alpha=0.15,0.20,0.25,0.30,0.35,0.40$ (plots from left to right). For $\alpha=0.15$ (first plot) we see usual decaying oscillations. For $\alpha=0.20$ two concurrent modes with the same damping rate. As $\alpha$ increases we observe exponentially damping tails, that do not oscillate. At higher $\alpha$ we see the oscillation behavior again, but the frequency of oscillation for $\alpha=0.40$ (last plot) differs significantly from that for $\alpha=0.15$.}
\end{figure*}

\begin{figure}
\includegraphics[width=.45\textwidth,clip]{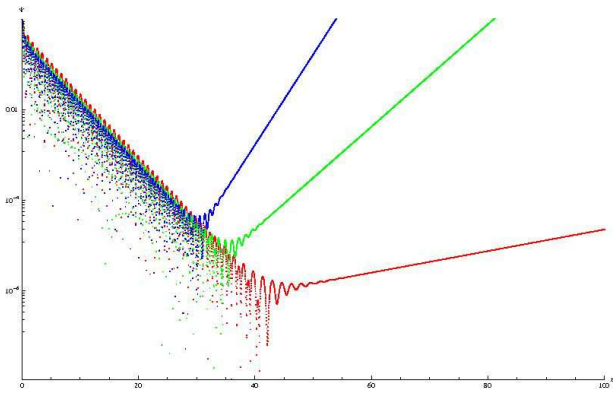}
\caption{The picture of instability, developing at large multipole
numbers: $D=6$, $\ell =8$ (red), $\ell = 12$ (green), $\ell=16$ (blue), $\alpha =1.3$. Tensor type of
gravitational perturbations.}
\end{figure}

\begin{figure}
\includegraphics[width=.45\textwidth,clip]{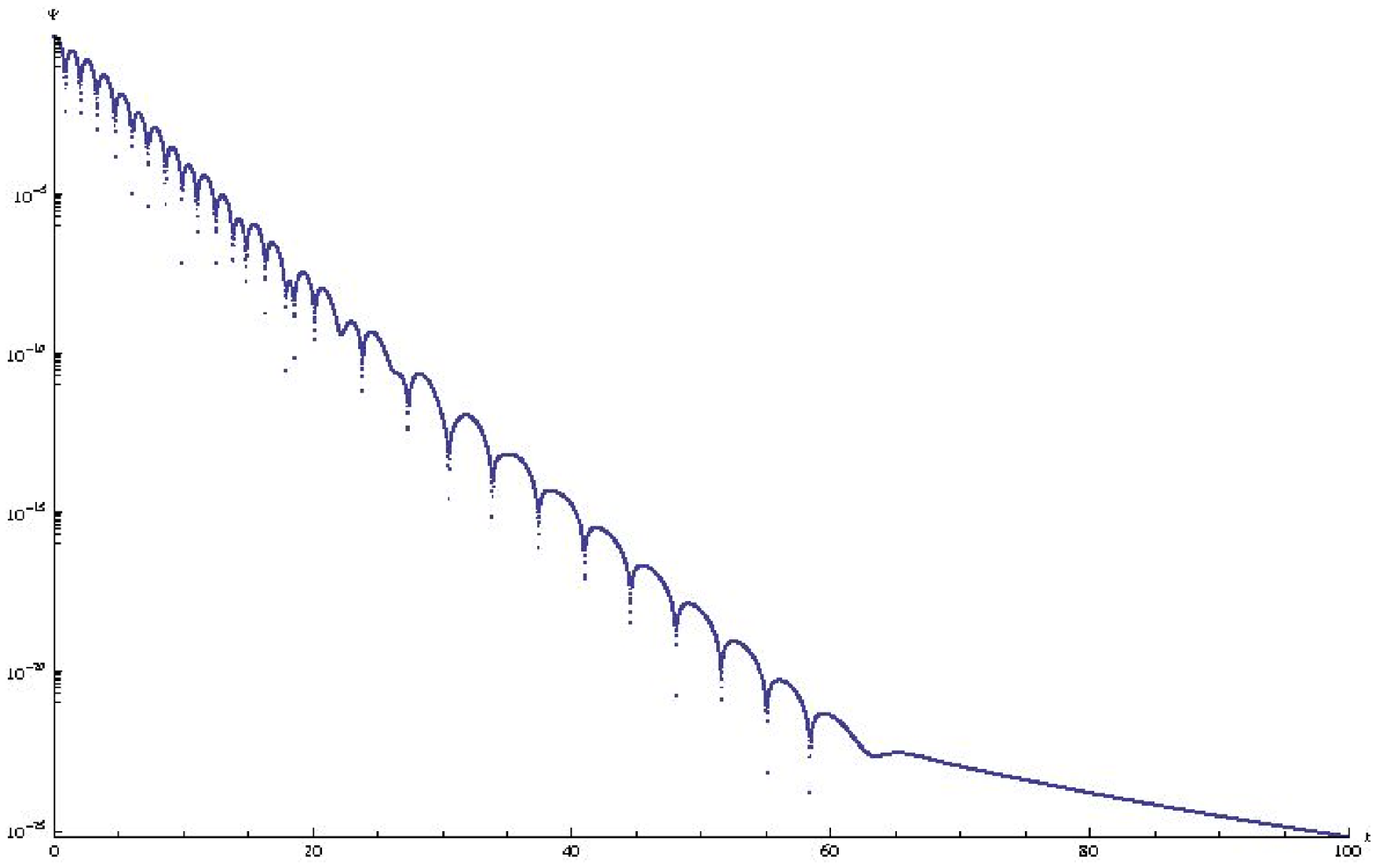}
\caption{The picture of time domain evolution for scalar type of
gravitational perturbations $D=10$, $\ell=2$, $\alpha=0.01$. One
can see that two modes are dominating at different stages, so that
in fact the signal is dominated by a superposition of the two
modes.}
\end{figure}

Let us note that for $D>4$ black holes in ordinary Einstein gravity
the tensor type of gravitational perturbations is governed by the
same wave equation as test scalar field. Therefore both types of
perturbations produce the same quasinormal mode spectrum, that is
they are isospectral. This coincidence is remarkable, but does not
take place for Gauss-Bonnet black holes. Indeed, if we compare the
tensor quasinormal modes in Table I in this paper with test scalar
field quasinormal modes of \cite{GBQNMsmy1}, \cite{GBQNMsmy2}, we
can see that quasinormal modes as well as effective potentials are
quite different.

Here we shall consider $\omega = \omega_{Re} - i \omega_{Im}$, and
the $\omega$ is chosen so that positive $\omega_{Im}$ corresponds
to a damped mode. On the tables I-III one can see fundamental
quasinormal modes obtained by time-domain integration. We see that
imaginary part of $\omega$, which is proportional to the damping
rate, is always decreasing, when $\alpha$ is increasing, for all
three types of gravitational perturbations and all $D$. In other
words the less $D$ and the stronger the Gauss-Bonnet coupling
$\alpha$, the slower decay of the perturbations is. In contrast
to the imaginary part, the real oscillation frequency $\omega_{Re}$
does not behave uniform: $\omega_{Re}$ decreases as
$\alpha$ grows for most cases of tensor and vector modes. The behavior of scalar
mode is different: here we have {\it two} competing for the domination
modes (see Fig. 6, for $D=10$) at different stages of the quasinormal
ringing. Thus for $D=7$, for example, at some values of $\alpha$
the two modes with close imaginary parts appear, what makes the
picture of time evolution more complicated. Then, at some larger
$\alpha$ the dominance takes another mode. This superposition of
modes, also with competing excitation coefficients of the particular
modes, makes dependence of the fundamental scalar type QNMs on
$\alpha$ and $D$ non-monotonic.

To check that our computation scheme is working properly and gives
no numerical error, we repeated the integration with much smaller
step and higher accuracy of all incoming data. The picture of the
evolution does not change, what means high stability and accuracy of
the integration. Another check may be going to particular known
limits, such as pure Schwarzschild case $\alpha =0$. Then we must
get pure Schwarzschild QNMs in that limit. Again the dominance of
the two modes at different stages of the ringing complicate the
picture. Indeed, for instance for

$D=10$ and $\alpha=0$ from \cite{MyNPB1} we have $\omega = 2.45-0.98
i$, what is not a fundamental mode for the whole stage of ringing
but rather for the first period. Indeed, we can see the approaching
of our GB QNMs to the pure Schwarzschild ones in the following data:

\begin{center}
\begin{tabular}{|c|c|c|}
  \hline
  $\alpha$  & $\omega_{0}$ & $\omega_{1}$ \\
  \hline
  0.00001  & 1.2347-0.9328 i & 2.4576-0.9873 i \\
  0.0001  & 1.2298-0.93 i  & 2.4580-0.9867 i  \\
  0.001  & 1.1834-0.9116 i & 2.4615-0.9807 i \\
  0.01  & 0.8960-0.7825 i  & 2.4387-0.9336 i \\
  0.1  & 0.8831-0.4418 i & 2.0611-0.8058 i  \\
  \hline
\end{tabular}
\end{center}

Therefore we can conclude that there is a kind of conceptual gap of
what one can consider as a fundamental mode of the quasinormal
ringing. Probably, the better choice would be to consider the last
stage of ringing, immediately before the tail stage as the one where
the fundamental quasinormal modes must be defined.

Another essential and distinctive feature of the Gauss-Bonnet
quasinormal ringing is that instability occurs at higher multipole
numbers $\ell$, while lowest $\ell$ are stable! This is indeed
remarkable as, naively, one would expect that if the lowest
multipole is stable, then higher multipoles just raise up the pick
of the potential barrier, so that higher multipoles should stabilize
the potential. Yet, for Gauss-Bonnet black holes, higher multipoles
also increase the negative gap near the black hole horizon
\cite{Dotti1}, \cite{Dotti2}, allowing existence of bound states in
the gap. This instability at higher multipoles seems to be intrinsic
to Gauss-Bonnet theories, because similar instability was found some
time ago in \cite{Soda-san} for Gauss-Bonnet cosmologies.

One can see the evolution of instability in time domain in Fig. 5.
The larger $\ell$, at the earlier times instability growth occurs,
and the stronger the growth rate is (Fig. 5). In addition, the
higher $\ell$, the smaller threshold value $\alpha$ at which
instability happens. At $\alpha$ smaller than the critical value, black
holes are stable. Therefore it is very important not to be limited
by small $\ell$ but to see the regime high multipoles, in order to
determine the threshold $\alpha$ with good accuracy. Analytical
estimations of \cite{Dotti2} are compared here with our numerical
results in Fig.1, 2. There one can see that the estimation of
\cite{Dotti2}, which gives an expression for the threshold value
$\alpha \approx A \ell^{-1} + B$, valid for not very large
$\ell$, while at larger $\ell$ instability occurs at smaller
$\alpha$. This decreases the minimal value of $\alpha$, when
Gauss-Bonnet black holes become unstable. Essential point is that
the above described instability exists only for $D=5$ and $6$, while
at higher $D$ the black holes are stable.

It should be explained here that in Table III we have two values of
$\alpha$ which do not correspond to instabilities, yet do not have
any definite quasinormal modes. The point is that for scalar type of
gravitational perturbations, in addition to the negative gap which
may deepen when $\ell$ is increasing, there is a negative gap at
higher $D$, which is less deep at higher $\ell$. This gap does not
produce instability, but may suppress the quasinormal ringing by a
exponential tail behavior, so that we almost do not see any ringing period (See
Fig.3, 4). For such a situation, it would be too strong to state
that the appearing very short period of oscillations with small real
frequency is dominated by some quasinormal modes. This ``transition''
period happens at intermediate values of $\alpha$, and then some
other mode start dominating. We stress that this rather odd
picture, in no way can be suspected as an instability, because this
"transition" behavior happen also for some parameters, for test
scalar field, when stability can be proved analytically. This kind
of negative gap at $D>5$, is similar to that for the
Reissner-Nordstr\"om black holes \cite{MyNPB1}, which is not deep
enough and do not produce the instability.

\section{Discussion}

In this paper the numerical integration in time domain was done for
the gravitational perturbations of black holes in $D=5-11$
Gauss-Bonnet theories. The instability occurs only for $D=5$ and
$D=6$ cases at some large values of $\alpha$. Higher $D$ stabilize
the perturbations. The instability starts after some period of
quasinormal ringing and at earlier time for larger multipoles
$\ell$. Apparently the instability at large $\alpha$ is a physically
expected result: Gauss-Bonnet black holes is inspired by a one-loop
string theory approximation, so that GB theory is valid only as soon
as $\alpha$ is small enough. Otherwise one needs to take into
consideration higher order corrections.

It is interesting to generalize the present work to the case of
charged Gauss-Bonnet black holes and asymptotically de-Sitter black
holes, because we know that $\Lambda$-term and the black hole charge
give also negative gap to the effective potential, so that if taking
into account all these factors, $\alpha$-coupling, and $\Lambda$ and
or charge, the parameters range of instability might be increased.

\begin{acknowledgments}
We would like to thank Prof. Elcio Abdalla for stimulating discussion and critical reading of the manuscript.\\
GNU Multiple Precision Arithmetic Library was used to obtain the presented results.\\
R. K. was supported by postdoctoral fellowship of
Japan Society for Promotion of Science (JSPS), and by Grant-in-Aid
of JSPS, Japan. R. K. also acknowledges hospitality of the
Department of Physics of Ioannina University (Greece), where the
final part of this work was done.\\
A. Z. thanks the EDGE Team for their work. A. Z. was supported by
\emph{Funda\c{c}\~ao de Amparo \`a Pesquisa do Estado de S\~ao Paulo
(FAPESP)}, Brazil.

\end{acknowledgments}

\end{document}